\documentclass[12pt]{article}
\usepackage{color} 
\usepackage{amsmath}
\usepackage{amssymb}
\usepackage{bm}
\usepackage[dvipdfm]{graphicx} 

\newcommand{\bea}{\begin{eqnarray}}
\newcommand{\eea}{\end{eqnarray}}
\newcommand{\be}{\begin{equation}}
\newcommand{\ee}{\end{equation}}

 \makeatletter  
\def\alt{\mathrel{\mathpalette\gl@align<}}
\def\agt{\mathrel{\mathpalette\gl@align>}}
\def\gl@align#1#2{ \lower.6ex\vbox{\baselineskip\z@skip\lineskip\z@
\ialign{ $\m@th#1\hfil##\hfil$\crcr#2\crcr\sim\crcr }} } \makeatother

\begin{document}
\begin{flushright}
KEK-TH-1600 \\ 
OIQP-13-01\\
HUPD-1211
\end{flushright}

%
\begin{center}

\baselineskip 20pt 
{\Large\bf 
On the Cancellation Mechanism of  Radiation \\ from the Unruh detector }
\vspace{1cm}

{
\large
 S. Iso$^{1,2}$\footnote{satoshi.iso@kek.jp},
 K. Yamamoto$^{3}$\footnote{kazuhiro@hiroshima-u.ac.jp},
 S. Zhang$^{4}$\footnote{lightondust@gmail.com}
}
 \vspace{.5cm}

{\baselineskip 20pt \it
$^{1}$ Theory Center, High Energy Accelerator Research Organization(KEK) \\
$^{2}$Department of Particles and Nuclear Physics, \\
 The Graduate University for Advanced Studies (SOKENDAI), \\
Oho 1-1, Tsukuba, Ibaraki 305-0801, Japan \vspace{2mm} \\
$^{3}$ Department of Physics, Hiroshima University, \\
         Higashi-Hiroshima 739-8526, Japan \vspace{2mm} \\
$^{4}$ Okayama Institute for Quantum Physics, \\
Kyoyama 1-9-1, Kita-ku, Okayama 700-0015, Japan
\vspace{3mm}
}

\vspace{2cm} 
\end{center}
\abstract{
A uniformly accelerated detector (Unruh detector) in the Minkowski vacuum is excited as if it 
is exposed to the thermal bath with temperature proportional to its acceleration.
In the inertial frame, since both of an excitation and a deexcitation of the detector 
are accompanied by emission of radiation into the Minkowski vacuum, 
one may suspect that the Unruh detector emits radiation like
the  Larmor radiation from an accelerated charged particle.
However, it is known that the radiation is miraculously cancelled 
by a quantum interference effect.
In this paper, we investigate under what condition the radiation cancels out.
We first show that the cancellation occurs if the Green function satisfies a  relation
similar to the Kubo-Martin-Schwinger (KMS) condition.
We then study two examples,  Unruh detectors in the 3+1 dimensional Minkowski spacetime
and in the de Sitter spacetime. 
In both cases, the relation
 holds only in a restricted region of the spacetime,
but the radiation is cancelled in the whole spacetime.
Hence the KMS-like relation
 is necessary but not  sufficient for the cancellation to occur.
}
\thispagestyle{empty}
\newpage

\addtocounter{page}{-1}

\baselineskip 18pt

\section{Introduction}
A uniformly accelerated observer sees the Minkowski vacuum as thermally excited, 
which is known as the Unruh effect \cite{Unruh:1992sw}. 
The Unruh effect is fundamental and important because it is 
related by the equivalence principle
to the thermal behavior of gravity in spacetime with horizons  \cite{HawkingRadiation}. 
The Unruh temperature $T_U=\hbar a/2\pi c k_B$ is proportional to 
its acceleration $a$ and very small for ordinary experimental settings.
But the recent developments of ultra-high intense lasers  make the 
Unruh effect experimentally accessible~\cite{ELI}. 
For example, Chen and Tajima~\cite{ChenTajima} proposed  an indirect detection
of the Unruh effect by measuring an excess of radiation from an 
accelerated electron in the electromagnetic field  of ultra-high intense lasers.
Since the trajectory of a  charged particle in acceleration fluctuates around 
the classical trajectory  due to the Unruh effect, 
it may emit an extra  radiation 
(Unruh radiation) besides the classical Larmor radiation.
The idea was further investigated in Ref.~\cite{Iso:2010yq}, 
in which it is shown that the motion of an
electron in  uniform acceleration is thermalized and fluctuates 
around the classical trajectory. 
An interference effect  plays an important role to cancel
the Unruh radiation at least partially, though it is not yet established whether
the radiation is totally canceled.

A similar analysis was investigated for a uniformly accelerated Unruh detector, i.e.
a system of a harmonic oscillator coupled with the quantum field. 
The backreaction of emission of particles 
to the detector's trajectory is neglected, so the trajectory of the detector 
is not a dynamical variable.
In \cite{Grove:1986,Massar:2005vg,Brout:1995rd,Sciama,Lin:2005uk},
it was shown that cancellation of  radiation
occurs due to an interference effect.
Here we briefly explain the mechanism of cancellation.
The system is described by  quantum scalar field $\phi$ and the Unruh detector in 
uniform acceleration.
The equation of motion of the scalar field is given by
\begin{eqnarray}
\partial^\mu \partial_\mu \phi(x) = j(x)
\end{eqnarray}
where the scalar current $j(x)$ is induced by the Unruh detector.
Its solution $\phi(x)$ is written as a sum of  an inhomogeneous  and a homogeneous solutions
\begin{eqnarray}
\phi(x) = \phi_{inh}(x) + \phi_{h}(x).
\end{eqnarray}
The inhomogeneous solution $\phi_{inh}(x)$ describes the field induced by the coupling to
the Unruh detector while the homogeneous part $\phi_{h}(x)$ 
corresponds to the vacuum fluctuation. 
The normal-ordered two-point function is given by the sum
\begin{eqnarray}
\langle \phi(x) \phi(y) \rangle - \langle \phi_h(x) \phi_h(y) \rangle
= \langle \phi_{inh}(x) \phi_{inh}(y) \rangle
+ \langle \phi_{inh}(x) \phi_h(y) \rangle +\langle \phi_h(x) \phi_{inh}(y) \rangle.
\label{R2p}
\end{eqnarray}
The first term $\langle \phi_{inh}(x) \phi_{inh}(y) \rangle$ 
is a classical contribution in the presence of the Unruh detector. 
In addition to it, we also have the interference terms
$\langle \phi_{inh}(x) \phi_h(y) \rangle +\langle \phi_h(x) \phi_{inh}(y) \rangle$.
It is purely quantum mechanical and appears
as a result of the interference between the 
induced field $\phi_{inh}$ and the vacuum fluctuation $\phi_h$.
Since the inhomogeneous solution has its origin in the vacuum fluctuation, 
the interference terms do not vanish.
It is indeed shown that these two contributions are  miraculously cancelled
each other (apart from the polarization cloud near the detector)  in some specific examples.  
 
Although the cancellation is straightforwardly shown, 
it is not clear under what condition the cancellation generally occurs.
An intuitive interpretation of the cancellation is based on an observation
that the Unruh detector  eventually reaches equilibrium 
in the thermal bath  with the Unruh temperature.
This may explain why the total energy
flux vanishes.
It sounds plausible, but it cannot answer to the following two questions. 
First, the uniformly accelerated observer can only 
see a part of the spacetime, the right Rindler wedge (see also section 3). 
So it is not certain whether the cancellation  occurs also in the future wedge
to which the accelerated observer is inaccessible.
Second, such classical interpretation of the cancellation cannot explain 
why the cancellation occurs for an inertial observer. 
In the thermal equilibrium, the energy balance is reached by  processes of absorption and 
emission. But, for an inertial observer, only emission can occur since there are no 
particles to be absorbed in the vacuum.
Both   absorption and   emission processes for the uniformly accelerated observer 
correspond to   emission of a quanta for the inertial observer.
Hence, the cancellation of the energy flux appears mysterious.

In this paper, we first investigate the mechanism of the cancellation
and find a  condition for the cancellation.
The condition is  similar to the Kubo-Martin-Schwinger (KMS) relation in a thermal system.
This makes clear and explicit the relation between the thermal properties of the uniformly accelerated 
observer and the cancellation of  radiation.
We then consider two examples, the Unruh detector in 3+1 dimensional Minkowski spacetime and a detector 
moving along a geodesic in de Sitter spacetime.
In both cases we show that a two-point function has different behaviors in  different wedges across the 
Rindler horizon of the  detector.
The above condition for the cancellation is only satisfied  in a 
restricted region of the spacetime (the right Rindler wedge).
This may reflect the fact that the uniformly accelerated observer can only access to the restricted
 part of the spacetime.  
In the future wedge, we show that, although the above thermal condition is not satisfied,
 radiation  also  cancels out and only a polarization cloud remains.

The paper is organized as follows. In Section 2, we give a general 
condition for the cancellation of radiation to occur.
In Section 3, we explicitly demonstrate how the condition 
works to cancel the radiation from an Unruh detector moving at a constant 
acceleration in the Minkowski spacetime. 
We also consider a detector at rest in the de Sitter spacetime.
Section 4 is devoted to conclusions and discussions. 

\section{KMS-like condition} 
\setcounter{equation}{0}

We consider a coupled system of a scalar field $\phi(x)$ and a harmonic oscillator
whose action is given by
\begin{eqnarray}
S[Q,\phi ; z] &=&
\frac{m}{2} \int d \tau \left( (\dot{Q}(\tau))^2 - \Omega_0^2 Q^2 \right) 
+ {1\over 2}\int d^n x \sqrt{|g|} \left(\partial^\mu \phi(x) \partial_\mu \phi(x) + F(R) \phi^2\right) \nonumber \\
& & + \lambda \int d^n x d\tau \ P[Q(\tau)] \phi(x) \delta^n(x-z(\tau)).
\label{SSS}
\end{eqnarray}
$Q(\tau)$ is
a harmonic oscillator with a mass $m$ and an angular frequency $\Omega_0$, and
denotes the dynamical degree of freedom of the Unruh detector.
Its world line trajectory is given by $x^\mu=z^\mu(\tau)$.
Note that $z^\mu(\tau)$ is not a dynamical variable.
$\phi(x)$ is  coupled to the Unruh detector though
the last term. $F(R)$ is a function of the Riemann scalar curvature.
$P[Q]$ is  defined by
\begin{align}
P[Q(\tau)] = \sum_{j} p_j \left( \frac{d}{d\tau} \right)^j Q(\tau),
\end{align}  
where $p_i$ is a constant.

The Heisenberg equations of motion are given by
\begin{eqnarray}
m \left( \ddot{Q}(\tau) + \Omega_0^2 Q(\tau) \right) &=& \lambda \bar{P}[\phi(z(\tau))], 
\label{EQQ}
\\
\left( \nabla^\mu \nabla_\mu \phi(x) - F(R) \right)\phi(x) 
&=& \frac{\lambda}{\sqrt{|g|} } \int d\tau' P[Q(\tau')] \delta^n(x-z(\tau')),
\label{EQphi}
\end{eqnarray}
where $\bar{P}[Q(\tau)] = \sum_{j} p_j \left(-\frac{d}{d\tau}\right)^j Q(\tau)$ is a conjugate  of $P$.
An inhomogeneous solution of (\ref{EQphi})  is given  by 
\begin{eqnarray}
\phi_{inh}(x) = \lambda \int d\tau' P[Q(\tau' )] G_R(x,z(\tau')),
\label{phiinh}
\end{eqnarray}
where the retarded Green function $G_R(x,y)$  satisfies
\begin{eqnarray}
\left( \nabla^\mu \nabla_\mu - F(R) \right) 
 G_R(x,y) = \frac{\delta^n(x-y)}{\sqrt{|g|} } 
\end{eqnarray}
with an appropriate boundary condition.
A general solution is then written as a sum of an inhomogeneous  and
a homogeneous solutions
\begin{eqnarray}
\phi(x) = \phi_{inh}(x) + \phi_h(x).
\label{GE}
\end{eqnarray}
Since we are considering an open 
system\footnote{If we consider a closed system (such as a system confined 
in a small box), we need to take into account a nonequilibrium evolution of the $\phi$ field 
by using the in-in formalism. Then the homogeneous solution deviates from the vacuum fluctuation.},
the homogeneous solution
 $\phi_h(x)$ describes the vacuum fluctuation and satisfies
\begin{eqnarray}
\left( \nabla^\mu \nabla_\mu - F(R) \right) \phi_h(x) = 0.
\end{eqnarray}

Substituting the solution~(\ref{GE}) into the equation of motion for $Q(\tau)$, Eq.~(\ref{EQQ}),
we obtain the following equation,
\begin{eqnarray}
m \left( \frac{d^2}{d\tau^2} + \Omega^2_0 \right) Q(\tau)
-\lambda^2 \bar{P} \left[ \int d\tau' P[Q(\tau')] G_R(z(\tau),z(\tau')) \right] 
= \lambda \bar{P} [\phi_h(z(\tau))].
\label{equationq0}
\end{eqnarray}
It describes a stochastic behavior of the Unruh detector, and 
the detector eventually reaches the thermal equilibrium at the Unruh temperature.
The second term in the l.h.s. gives a dissipation due to the emission of radiation
while the r.h.s. gives a stochastic  noise of the quantum vacuum fluctuation. 
In the following, we will consider a trajectory of the detector so
that $G_R(z(\tau),z(\tau'))$ is a function of $\tau-\tau'$,
$$G_R(z(\tau),z(\tau')) = G_R(\tau-\tau').$$
This condition  holds  in two examples studied in section 3.
After the Unruh detector reaches the thermal equilibrium, 
the coupled system becomes  stationary (but not necessarily static).
We then  Fourier-transform $Q(\tau)$ as
\begin{eqnarray}
Q(\tau) &=& \int_{-\infty}^\infty \frac{d\omega}{2\pi} e^{-i\omega \tau} \tilde{Q}(\omega).
\end{eqnarray}
Similarly the vacuum fluctuation $\phi_h(z(\tau))$ along the trajectory of the detector
can be Fourier transformed as
\begin{eqnarray}
\tilde{\phi}(\omega) = \int d\tau \ e^{i\omega \tau} \phi_h(z(\tau)).
\end{eqnarray}
Then the equation (\ref{equationq0}) can be  solved in terms of $\tilde{\phi}(\omega)$ as
\begin{eqnarray}
\tilde{Q}(\omega) = \lambda h(\omega) \tilde{\phi}(\omega).
\label{Qsolution}
\end{eqnarray}
Here $h(\omega)$ is given by
\begin{eqnarray}
h(\omega) = \frac{f(-\omega)}{-m\omega^2 + m\Omega_0^2 - \lambda^2 f(\omega) f(-\omega) \tilde{G_R}(\omega)},
\label{homega}
\end{eqnarray}
where we defined $f(\omega) = \sum_j p_j (-i\omega)^j$, and 
\begin{eqnarray}
\tilde{G}_R(\omega) = \int d(\tau-\tau') G_R(\tau-\tau') e^{i\omega (\tau-\tau')} 
= \tilde{G}^{*}_R(-\omega).
\end{eqnarray}
Note that the relation $h(-\omega) = h^*(\omega)$ holds.

\vspace{5mm}
Now we consider the renormalized two-point function Eq.~(\ref{R2p}). 
The energy-momentum flux can be obtained from the two-point function by differentiating it with respect to $x$ and $y$.
Hence we focus our investigation on the two-point function. 

Using  (\ref{Qsolution}), the inhomogeneous solution  $\phi_{inh}$ (\ref{phiinh}) 
can be written in terms of the homogeneous one (vacuum fluctuation)
$\tilde{\phi}(\omega)$ as
\begin{eqnarray}
\phi_{inh}(x) = \lambda^2 \int d \tau_x \int \frac{d\omega_x}{2\pi} f(\omega_x)  h(\omega_x) \tilde{\phi}(\omega_x)
\ G_R(x,z(\tau_x )) e^{-i\omega_x\tau_x}.
\end{eqnarray}
Then the two-point correlation of the inhomogeneous solution is given by
\begin{eqnarray}
\langle \phi_{inh}(x) \phi_{inh}(y) \rangle 
&=& \lambda^4 \int \frac{d\tau_x d\omega_x d\tau_y d\omega_y}{(2\pi)^2} \ 
G_R(x,z(\tau_x )) G_R(y,z(\tau_y )) e^{-i(\omega_x \tau_x + \omega_y \tau_y)} \nonumber \\
& & \hspace{20mm} \times \ h(\omega_x) f(\omega_x) h(\omega_y) f(\omega_y) 
\langle \tilde{\phi}(\omega_x) \tilde{\phi}(\omega_y)\rangle.
\label{inhinh}
\end{eqnarray}
Since the inhomogeneous solution $\phi_{inh}$ is written in terms of the homogeneous one $\phi_h$,
the interference between them gives a nonvanishing contribution to the two-point function.
It is given by
\begin{eqnarray}
& & \langle \phi_{inh}(x) \phi_{h}(y) \rangle + \langle \phi_{h}(x) \phi_{inh}(y) \rangle \nonumber \\
&=& \lambda \int d\tau_x G_R(x,z(\tau_x)) \langle P[Q(\tau_x)] \phi_h(y) \rangle 
+ \lambda \int d\tau_y G_R(y,z(\tau_y )) \langle \phi_h(x) P[Q(\tau_y)] \rangle   \nonumber \\
&=& \lambda^2 \int \frac{d\tau_x d\omega_x}{2\pi} G_R(x,z(\tau_x)) 
e^{-i\omega_x \tau_x} h(\omega_x) f(\omega_x) 
\langle \tilde{\phi}(\omega_x) \phi_h(y)\rangle \nonumber \\
& & + \lambda^2 \int \frac{d\tau_y d\omega_y}{2\pi} G_R(y,z(\tau_y)) 
e^{-i\omega_y \tau_y} h(\omega_y) f(\omega_y) 
\langle \phi_h(x) \tilde{\phi}(\omega_y) \rangle.
\label{inhh}
\end{eqnarray}
Comparing (\ref{inhinh}) and (\ref{inhh}),
one can see that (\ref{inhinh}) is written in terms of the correlation of vacuum fluctuations
on the trajectory $\langle \phi_h(z(\tau))\phi_h(z(\tau')) \rangle$  while (\ref{inhh}) depends on $\langle \phi_h(x)\phi_h(z(\tau)) \rangle$, and  
 a  nontrivial relation is necessary to make them related.
 
In the remaining of this section, we show that the following  relation
plays an important role to cancel out the radiation.
The key relation we will use is 
\begin{eqnarray}
\langle \tilde{\phi}(\omega) \phi_h(y) \rangle = 
\rho(\omega) \langle [\tilde{\phi}(\omega),\phi_h(y)] \rangle,
\label{iyz}
\end{eqnarray}
where $\rho(\omega)$ is a real function of $\omega$. 
Or equivalently,
\begin{align}
\langle \tilde{\phi}(\omega) \phi_h(y) \rangle = \alpha(\omega) \langle \phi_h(y) \tilde{\phi}(\omega) \rangle, \ \ \ 
\alpha(\omega) = {{\rho(\omega)}\over {\rho(\omega)}-1}.
\end{align}
These relations show that 
$\langle \tilde{\phi}(\omega) \phi_h(y) \rangle$
and  
$\langle [\tilde{\phi}(\omega),\phi_h(y)] \rangle$
have the same $y$-dependence up to a real function of $\omega$.
Both of them satisfy the homogeneous equation 
$\left( \nabla^\mu \nabla_\mu - F(R) \right) G(y) = 0 $, but 
it does not mean that the relation (\ref{iyz}) is always satisfied.

We now prove that the radiation in the 
inhomogeneous term (\ref{inhinh}) 
and  the interference term  (\ref{inhh}) cancels out
when the key relation Eq.~(\ref{iyz}) holds. 
Introducing a function $G(x,y)$ by
\begin{eqnarray}
G(x,y) = - i \langle [\phi_h(x),\phi_h(y)] \rangle = - G_R(x,y) + G_A(x,y), 
\end{eqnarray}
(see e.g., \cite{BirrelDavies}), we find that
\begin{eqnarray}
\langle \tilde{\phi}(\omega_x) \tilde{\phi}(\omega_y) \rangle &=&
\int d\tau_y \ \langle \tilde{\phi}(\omega_x) \phi_h(z(\tau_y)) \rangle e^{i\omega_y \tau_y}
\nonumber \\
&=& i \int d\tau_x d\tau_y \ \rho(\omega_x) G(\tau_x - \tau_y) 
e^{i(\omega_x \tau_x + \omega_y \tau_y)} \nonumber \\
&=& 4\pi \delta(\omega_x + \omega_y) \rho(\omega_x) \ \mbox{Im} \tilde{G}_R(\omega_x).
\label{phiinhphiinh}
\end{eqnarray}
In the last equality  we used the relation $G_A(x,y) = G_R(y,x)$ and   
$\tilde{G}(\omega) = -\tilde{G}_R(\omega) + \tilde{G}_A(\omega) = -2i \ \mbox{Im} \tilde{G}_R(\omega)$,
where $\tilde G(\omega)$ and $\tilde G_{A}(\omega)$ are the Fourier transforms of 
$G(z(\tau),z(\tau'))=G(\tau-\tau')$ and $G_A(\tau(z),\tau(z'))=G_A(\tau-\tau')$, respectively.
Substituting this relation into Eq.~(\ref{inhinh}), we have
\begin{eqnarray}
& &\langle \phi_{inh}(x) \phi_{inh}(y) \rangle \nonumber \\
&&=\lambda^4 \int \frac{d\tau_x d\tau_y d\omega}{2\pi} 
G_R(x,z(\tau_x)) G_R(y,z(\tau_y)) |h(\omega)f(\omega)|^2  
\rho(\omega)  e^{- i\omega (\tau_x - \tau_y)} 2 \ \mbox{Im} \tilde{G}_R(\omega).
\nonumber\\
\label{inhinh2}
\end{eqnarray}
On the other hand, by using the key relation (\ref{iyz})
the interference term (\ref{inhh}) becomes
\begin{eqnarray}
&& \langle \phi_{inh}(x) \phi_{h}(y) \rangle + \langle \phi_{h}(x) \phi_{inh}(y) \rangle \nonumber \\
&&= -i \lambda^2 \int \frac{d\tau_x d\tau_y d\omega}{2\pi} 
e^{-i\omega(\tau_x-\tau_y)} \rho(\omega) 
\nonumber \\
&& \hspace{2mm}\times
\left(
G_R(x,z(\tau_x)) G(y,z(\tau_y)) f(\omega) h(\omega)  - 
G(x,z(\tau_x)) G_R(y,z(\tau_y)) f(-\omega) h(-\omega) \right),
\nonumber\\
\label{inhh2}
\end{eqnarray}
 where we used 
\begin{eqnarray}
\langle \phi_h(x) \tilde{\phi}(\omega) \rangle = (\langle \phi_h(x) \tilde{\phi}(\omega) \rangle^*)^* =
(\langle \tilde{\phi}(-\omega) \phi_h(x) \rangle)^* = \rho(-\omega) \int d\tau (-i) G(z(\tau),x) e^{i\omega \tau}.
\nonumber\\
\end{eqnarray}
Since the commutator $G(x,y)$ is written as 
 $G(x,y) = G_A(x,y) - G_R(x,y)$, (\ref{inhh2}) can be decomposed into
a term containing a product of two $G_R$ and the other with a product of $G_R$ and $G_A$.
It can be easily shown by using the identity
\begin{eqnarray}
h(\omega) f(\omega) - h(-\omega)f(-\omega) 
&=& { \lambda^2}|h(\omega)|^2 
\left( |f(\omega)|^2 \tilde{G}_R(\omega) - |f(\omega)|^2 \tilde{G}_R(-\omega) \right) \nonumber \\
&=& |h(\omega) f(\omega)|^2 2i \lambda^2 \ \mbox{Im} \tilde{G}_R(\omega)
\end{eqnarray}
that the term containing a product of two $G_R$ in (\ref{inhh2}) totally cancels
the two-point function
$\langle \phi_{inh}(x) \phi_{inh}(y) \rangle$ in  (\ref{inhinh2}),
which would generate the classical radia1tion by the fluctuating motion of the 
accelerated detector.

As a result of the above cancellation, 
the  renormalized two-point function becomes
\begin{eqnarray}
& & \langle \phi(x) \phi(y) - \phi_h(x) \phi_h(y) \rangle \nonumber \\
&=& -i\lambda^2 \int \frac{d\tau_x d\tau_y d\omega}{2\pi} e^{-i\omega(\tau_x-\tau_y)} 
\rho(\omega) 
 \label{polarization} \\  
& &  \times \left\{ G_R(x,z(\tau_x))G_A(y,z(\tau_y)) f(\omega)h(\omega) -
 G_A(x,z(\tau_x)) G_R(y,z(\tau_y)) f(-\omega)h(-\omega) \right\} . \nonumber 
\end{eqnarray}
It contains a product of $G_R$ and $G_A$, and because of this we can show that 
the energy-momentum tensor derived from this two-point function
damps faster than the behavior of  radiation. Hence it does not give an energy flux at infinity.
 We will see this explicitly in the next section. 
\section{Two Examples} 
\setcounter{equation}{0}
In this section we consider two examples to investigate the
mechanism of cancellation of radiation.
We will see that 
the key relation  (\ref{iyz}) holds only in a restricted region of the spacetime.
The first example is the Unruh detector that is
uniformly accelerated in the 3+1 dimensional Minkowski spacetime.
The second one is a  detector fixed at 
the origin of the spatial coordinates in the 3+1 de Sitter spacetime.

\subsection{Unruh detector in Minkowski spacetime}

We consider a massless scalar field  (\ref{SSS}) in the Minkowski spacetime .
For simplicity, we take $P[Q(\tau)]=Q(\tau)$ so that $f(\omega)=1$.
The vacuum two-point function 
$\langle  \phi(x) \phi(y) \rangle$ is a function of 
the invariant distance $\sigma = (x-y)^2$. 
The trajectory of a uniformly accelerated detector is given by
\begin{eqnarray}
z^\mu(\tau) = \left(\frac{\sinh{a\tau}}{a}, \frac{\cosh{a\tau}}{a},0,0\right),
\end{eqnarray}
and the invariant distance between  two points, $z^\mu(\tau)$ and $z^\mu(\tau')$,
on the trajectory is given by 
\begin{eqnarray}
\sigma = (z(\tau)-z(\tau'))^2=
\frac{4}{a^2} \left(\sinh{\frac{a(\tau-\tau')}{2}} \right)^2.
\end{eqnarray}
It is a function of the difference of the detector's proper time, $\tau-\tau'$.
Therefore, the Green function is a function of $(\tau-\tau')$;
 $G_R(z(\tau),z(\tau'))=G_R(\tau - \tau')$.

In 3+1 dimensional Minkowski spacetime, the Wightman function is given by
\begin{eqnarray}
\langle \phi_h(x) \phi_h(y) \rangle = - \frac{1}{4\pi^2} \frac{1}{(x-y)^2 - i\epsilon (x^0-y^0)}
\end{eqnarray}
where $\epsilon$ is an infinitesimally small positive constant.
To explore when the key relation~(\ref{iyz}) holds, we calculate the following quantity,
\begin{eqnarray}
\langle \phi_h(x) \tilde{\phi}(\omega) \rangle &=& 
\int d\tau e^{i\omega \tau} \langle \phi_h(x) \phi_h(z(\tau)) \rangle \nonumber \\ 
&=& - \frac{1}{4\pi^2} \int d\tau 
   \frac{e^{i\omega \tau} }
	     {(x^0 -z^0(\tau) -i\epsilon)^2 -(x^1 -z^1(\tau))^2 -(x^2)^2 -(x^3)^2 }.
\end{eqnarray}
The  integrand has poles on the complex $\tau$ plane, whose
positions are obtained by solving the equation
\begin{eqnarray}
    \left(x^0 -\frac{\sinh(a\tau)}{a} \right)^2 
  - \left(x^1 -\frac{\cosh(a\tau)}{a} \right)^2 - (x^2)^2 - (x^3)^2=0.
\end{eqnarray}
In terms of the lightcone coordinates  $u=x^0-x^1$ and $v=x^0+x^1$,
it becomes 
\begin{eqnarray}
  - u \frac{e^{a\tau}}{a} + v \frac{e^{-a\tau}}{a} 
  + x^\mu x_\mu -\frac{1}{a^2}=0.
\label{polelight}
\end{eqnarray}
The solutions of (\ref{polelight}) are obtained in a different form
depending on a given spacetime point.  We consider two types of observers,
one in the future wedge where  $u>0$ and $v>0$ and the other in the right
Rindler wedge where $u<0$ and $v>0$.

If $x$ is in the right Rindler wedge with $u<0$ and $v>0$,
there are two types of solutions and each of them
 satisfies the following equation,
\begin{align}
  e^{a\tau_-^R} &= 
    \frac{a}{2|u|}
    \Bigl( L^2 -\sqrt{ L^4 -\tfrac{4}{a^2}|uv| } \Bigr) >0 \\ 
   e^{a\tau_+^R} &= 
    \frac{a}{2|u|}
    \Bigl( L^2 +\sqrt{ L^4 -\tfrac{4}{a^2}|uv| } \Bigr) >0,
\end{align}
where $L^2 = -x^\mu x_\mu +1/a^2$.
Due to the thermal property of the accelerated observer, the solutions 
are periodically located at $\tau_\pm^R=\zeta_{\pm}^R + 2\pi n i/a$, where 
$\zeta_{\pm}^R$ are real-valued and defined by 
\begin{align}
    \zeta_\pm^R(x)={1\over a}\ln\left[
    \frac{a}{2|u|}
    \Bigl(L^2 \pm\sqrt{ L^4 -\tfrac{4}{a^2}|uv| } \Bigr)\right],
\end{align}
respectively.
On the other hand, 
if $x^\mu$ is in the future wedge with $u>0$ and $v>0$, the solutions satisfy
\begin{eqnarray}
  &&e^{a\tau_-^F} =     \frac{a}{2u}
    \Bigl( -L^2 +\sqrt{ L^4 +\tfrac{4}{a^2} uv } \Bigr) >0 \\ 
  &&e^{a\tau_+^F}=     \frac{a}{2u}
    \Bigl( -L^2 - \sqrt{ L^4 +\tfrac{4}{a^2} uv } \Bigr)<0.
\label{minussign}
\end{eqnarray}
The solutions are located periodically on the complex $\tau$ plane
as $\tau_-^F=\zeta_-^F + 2\pi n i/a$ and $\tau_+^F=\zeta_+^F + \pi (2n+1) i/a$, 
where real-valued $\zeta_{\pm}^F$ are defined by
\begin{align}
    \zeta_\pm^F(x)={1\over a}\ln\left[
    \frac{a}{2|u|}
    \Bigl(\pm L^2 +\sqrt{ L^4 -\tfrac{4}{a^2}|uv| } \Bigr)\right],
\end{align}
respectively.
Note that the imaginary parts of $\tau_+^F$ are different by $\pi/a$ from the other poles.
Since the imaginary parts of $\tau_+^F$ are half integers divided by $a$,
it is not a proper time of the detector's trajectory.
Rather one can interpret it as the proper time of a virtual trajectory 
in the left Rindler wedge.
(For further details, see Figure 2 in \cite{Iso:2010yq} .)

Summing these contributions to the integration, we obtain
\begin{eqnarray}
\langle \phi_h(x) \tilde{\phi}(\omega) \rangle =  
\frac{i}{4\pi l(x)} \frac{1}{e^{2\pi \omega/a}-1}
	 \bigl( 
	   e^{i\omega \zeta_-(x)} -e^{i\omega \zeta_{+}(x)} Z(\omega,x)  
	 \bigr),
\label{phitphi}
\end{eqnarray}
where 
\begin{eqnarray}
  Z_x &=& e^{\pi \omega /a} \theta (u) + \theta (-u) , \\ 
  l(x) &=& \dot{z}(\zeta_-)\cdot (x-z(\zeta_-))= \sqrt{ \tfrac{a^2}{4} L^4 + uv },
\label{lx}
\end{eqnarray}
Since the two-point function has different behaviors 
in the right Rindler wedge ($u<0$) and in the future wedge ($u<0$), 
we treat the them separately in the following.

\subsubsection{Right Rindler wedge}

In the right Rindler wedge  with $u<0$ and $v>0$, it is easy to verify that (\ref{iyz}) does hold
\begin{eqnarray}
\langle \tilde{\phi}(\omega) \phi_h(x) \rangle = ( \langle \phi_h(x) \tilde\phi(-\omega) \rangle )^* = e^{2\pi \omega/a}  \langle \phi_h(x) \tilde\phi(\omega) \rangle.
\end{eqnarray}
Hence $\rho(\omega) = 1/(1-e^{-2\pi \omega/a})$.
The right Rindler wedge is a region accessible by the accelerated observer.
For the accelerated observer, the Minkowski vacuum is seen as a thermal bath which makes 
the Unruh detector in thermal equilibrium. 
Therefore the key relation and accordingly the 
 cancellation of  radiation is physically understandable
in terms of the thermal behavior of the Unruh detector.

Let us now  show that the remaining term in the two-point function 
(\ref{polarization}) damps  faster than the behavior expected for  radiation so that
  it describes a polarization cloud around the detector.
The integral over $\tau_x$ and $\tau_y$ in (\ref{polarization}) 
can be performed by using the following  identities, 
\begin{eqnarray}
\int d\tau G_R(x-z(\tau)) q(\tau) &=& \frac{1}{4\pi l(x)} q(\tau_-), \nonumber \\
\int d\tau G_A(x-z(\tau)) q(\tau) &=& \frac{1}{4\pi l(x)} q(\tau_+),
\end{eqnarray}
where $q(\tau)$ is an arbitrary function and $l(x)$ defined in (\ref{lx})
is the distance measured by the comoving observer between  $x$ and $z(\tau_-)$.
Then,  (\ref{polarization}) becomes
\begin{eqnarray}
& & \langle \phi(x) \phi(y) - \phi_h(x) \phi_h(y) \rangle \nonumber \\
&=& -i\lambda^2 \int \frac{d\omega}{2\pi} \frac{\rho(\omega)}{(4\pi)^2 l(x) l(y)}
\left\{
e^{-i\omega(\zeta^R_- (x)- \zeta^R_+(y))} h(\omega) -
e^{-i\omega(\zeta^R_+(x) - \zeta^R_-(y))} h(-\omega) \right\} . \nonumber \\
\label{remaining}
\end{eqnarray}
The integral over $\omega$ can be evaluated by summing the residues of 
the poles of the functions $h(\omega)$ and $\rho(\omega)$.
The function $h(\omega)$ is given in (\ref{homega}), setting $f(\omega)=1$.
Since the retarded Green function in 3+1 dimensions is given by
$$G_R(x) = \frac{\theta(x^0) \delta(x^\mu x_\mu)}{2\pi},$$
$\tilde{G}_R(\omega)$ becomes
\begin{eqnarray}
\tilde{G}_R(\omega) = \int d\tau e^{i\omega \tau} \frac{\delta((z(\tau)-z(\tau'))^2)}{2\pi}
= \int d\tau e^{i\omega \tau} \frac{\delta(\tau-\tau')}{4\pi |\tau-\tau'|}.
\end{eqnarray}
The divergence in the real part of $\tilde{G}_R(\omega)$ gives 
a renormalization of  $\Omega_0$.
We write the renormalized frequency as  $\Omega$.
The imaginary part is given by $\omega/4\pi$, and we have
\begin{eqnarray}
h(\omega) = \frac{1}{-m\omega^2 + m\Omega^2 - i\frac{\omega \lambda^2}{4\pi}}.
\end{eqnarray}
The positions of the poles of $h(\omega)$ are hence given by  
\begin{eqnarray}
\omega_{\pm} = -\frac{i \lambda^2}{8\pi m} \pm \sqrt{\Omega^2 - \frac{\lambda^4}{64 m^2 \pi^2}}.
\end{eqnarray}
Since both of the poles $\omega_\pm$ are located on the lower complex plane of $\omega$, 
their contributions to the integral (\ref{polarization}) become proportional to 
$\theta(\zeta_-(x) - \zeta_+(y))$ or $\theta(\zeta_-(y) - \zeta_+(x))$, which vanish
when two points  $x,y$ coincide. Hence they do not give any contributions to the energy momentum tensor.
The poles of $\rho(\omega)=1/(1-e^{-2\pi\omega/a})$ are  located at
$\omega_{\pm n} = \pm na i$ with  a positive integer $n$. 
The pole at $\omega=0$ doesn't give any contribution because the residue vanishes.
The pole at $\omega_{\pm n}$ ($\omega \neq 0$) gives a term proportional to
$e^{-an|\zeta_-(x) - \zeta_+(y)|}$ or $e^{-an|\zeta_-(y) - \zeta_+(x)|}$ and it damps quickly at infinity. 
Indeed, we have
\begin{eqnarray}
e^{-a|\zeta^R_- - \zeta^R_+|} 
&=& - \frac{a^2}{4uv} \left( L^2 - \sqrt{L^4+\frac{4}{a^2}uv} \right)^2 \nonumber \\
&=& - \frac{a^2}{4uv} \left( \frac{2}{a}l(x) - \frac{2}{a} \sqrt{ l(x)^2 - uv } \right)^2
\longrightarrow - \frac{uv}{l(x)^2},
\end{eqnarray}
which damps  faster than $l(x)^{-1}$ at the infinity $l(x) \rightarrow \infty$. 
Together with $l(x)l(y)$ in the denominator of  (\ref{remaining}),
the two-point function damps faster than
  radiation which should behave as $\sim l(x)^{-2}$.
Hence there is no radiation in the right Rindler wedge.

\subsubsection{Future wedge}

In the future wedge with $u<0$ and $v>0$,  we have
\begin{eqnarray}
\langle \tilde{\phi}(\omega) \phi_h(x) \rangle 
&=& ( \langle \phi_h(x) \tilde\phi(-\omega) \rangle )^* 
= \frac{i}{4\pi l(x) } \frac{e^{2\pi\omega/a}}{e^{2\pi \omega/a}-1}
	 \bigl( 
	   e^{i\omega \zeta_-^F} -e^{i\omega \zeta^F_{+}-2\pi\omega/a}  
	 \bigr) \nonumber \\
&=& e^{2\pi \omega/a}  \langle \phi_h(x) \tilde\phi(\omega) \rangle + 
\frac{i}{4\pi l(x)} e^{i\omega \zeta^F_+ + \pi \omega/a}
\end{eqnarray}
and 
\begin{eqnarray}
\langle \tilde{\phi}(\omega) \phi_h(x) \rangle 
= \frac{1}{1-e^{-2\pi\omega/a}} \langle [\tilde{\phi}(\omega), \phi_h(x)] \rangle 
- \frac{i}{4\pi l(x)} \frac{e^{\pi\omega/a}}{e^{2\pi\omega/a}-1} e^{i\omega \zeta^F_+(x)}.
\label{FW}
\end{eqnarray}
The key relation (\ref{iyz}) does not hold exactly because of an additional term involving $\zeta^F_+$.
 Nevertheless we will show that there is no radiation in the future wedge. 
The first term in the r.h.s of Eq.~(\ref{FW}) gives 
the same contribution to the two-point function (\ref{polarization})
as discussed  in the right Rindler wedge, but since  $G_A(x-z(\tau))=0$ for $x$  in the future wedge, 
the contribution vanishes in this case.
So only the additional term in (\ref{FW})
gives a nonvanishing contribution to the  two-point function and we have
\begin{eqnarray}
&& \langle \phi(x) \phi(y) - \phi_h(x) \phi_h(y) \rangle \nonumber \\
&&=  \frac{i \lambda^2}{(4\pi)^2 l(x)l(y)} 
\int d\omega  \frac{e^{\pi\omega/a}}{e^{2\pi\omega/a}-1}
\left\{
h(\omega) e^{-i\omega(\zeta^F_- (x) - \zeta^F_+(y))} -
h(- \omega) e^{-i\omega(\zeta^F_+ (x) - \zeta^F_-(y))}   \right\} . \nonumber \\
\label{futureresult}
\end{eqnarray}
Although the key relation (\ref{iyz}) does not hold, the final result
(\ref{futureresult}) has a similar form to (\ref{remaining}).
In the future wedge, we have the relation
\begin{eqnarray}
e^{-a|\zeta^F_- - \zeta^F_+|} 
&=& \frac{a^2}{4uv} \left( L^2 - \sqrt{L^4+\frac{4}{a^2}uv} \right)^2 \nonumber \\
&=& \frac{a^2}{4uv} \left( \frac{2}{a}l(x) - \frac{2}{a} \sqrt{ l(x)^2 - uv } \right)^2
\longrightarrow \frac{uv}{l(x)^2}.
\end{eqnarray}
Hence the two-point function  damps as $l(x)^{-3}$ at the infinity $l(x)\rightarrow \infty$, 
and there is no radiation in the future wedge either.

In the future wedge, the key relation which reflects the thermal behavior of the Unruh detector
does not hold,
but the cancellation of the radiation still holds.
It is interesting that the additional term in 
(\ref{FW}) gives a 
contribution to the two-point function
which is similar
to the thermal contribution in the 
right Rindler wedge.
The remaining term describes a polarization cloud induced by the presence of
the accelerated Unruh detector.

\subsection{Detector at rest in de Sitter spacetime}
De Sitter spacetime is the maximally symmetric curved spacetime and the 
quantum field theory in de Sitter spacetime is the key to understand the 
early evolution of the universe. 
It is known that the quantum field theory in the de Sitter spacetime exhibits 
a similar feature as the Rindler noise. Namely, a detector at rest in 
de Sitter spacetime 
sees the Bunch-Davies vacuum as a thermally excited
state with the Gibbons-Hawking temperature. In this subsection, we explicitly 
show that the radiation from the detector cancels out due to the interference
effect and that the same theoretical structure reappears as in the Unruh detector 
in the Minkowski spacetime.

The de Sitter spacetime with a flat spatial chart is given by the line element,
\begin{eqnarray}
  ds^2=dt^2-a^2(t)d{\bf x}^2,
\end{eqnarray}
where $a(t) = e^{Ht}$ is the scale factor and $H$ is a constant. 
We consider a detector defined in Eq.~(\ref{SSS}), 
where it is fixed at the origin of the spatial coordinate, 
and its trajectory is defined by
\begin{eqnarray}
z^\mu(\tau) = (\tau,0,0,0).
\end{eqnarray}
Hence the proper-time of the detector is the same as the coordinate time $t$.
We consider a real scalar field with the conformal coupling to the curvature 
and set $F(R)=-{R}/{6}$ in Eq.~(\ref{SSS}).
\def\t{\tau}

\begin{figure}[htbp]
\begin{center}
\includegraphics[width=0.45\hsize,clip,angle=-90]{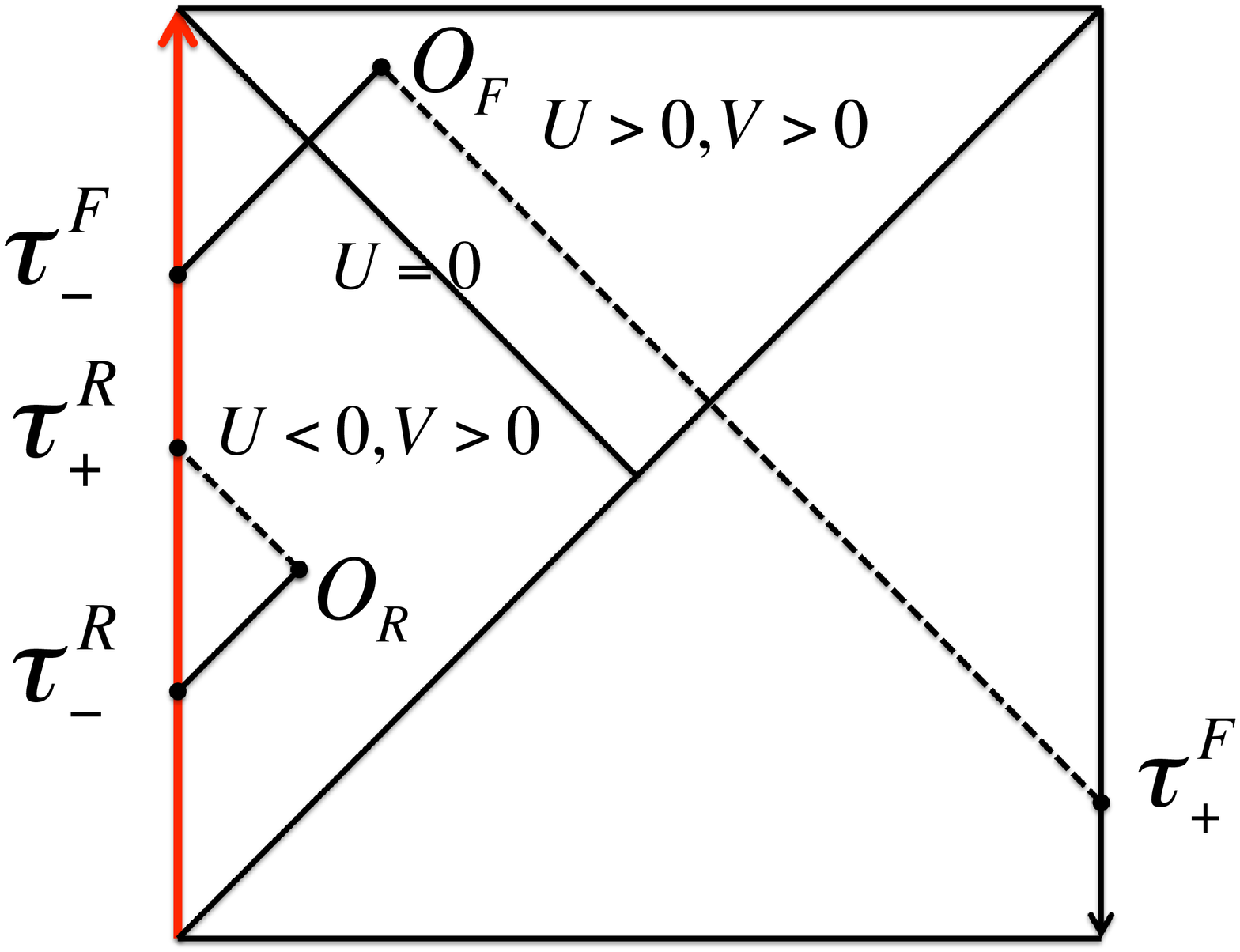}
\caption{A sketch of the conformal diagram of the de Sitter spacetime. 
The upper left region is covered by the coordinates of (\ref{deline})
with $-\infty<\eta<0$, while the lower right region is covered 
by similar coordinates with $0<\eta<\infty$ (e.g., \cite{HE}). 
In this diagram, the trajectory of the detector is located at the (red) left vertical 
axis.
The null surface $U=0$ divides the upper left region into 
the region with $U<0$ and $V>0$ (corresponding to the right Rindler wedge) and 
the region with $U>0$ and $V>0$ (corresponding to the future wedge). 
The points denoted by $\tau_-^R$ and $\tau_+^R$ correspond to the positions of
the solutions (\ref{detR}). Similarly, $\tau_-^F$ 
and $\tau_+^F$ show the position of the solution of 
(\ref{detF}), but because of the minus sign of the r.h.s. of (\ref{eht})
in the region $U>0$ and $V>0$, the position of 
$\tau_+^F$ lies on a virtual trajectory lying in the extended spacetime 
region with a positive conformal time $\eta>0$.
}
\label{fig:figds}
\end{center}
\end{figure}

Similar to the case of the Unruh detector in the Minkowski spacetime, 
we start from evaluating the correlator $\langle \phi_h(x) \tilde{\phi}(\omega) \rangle$.
To this end, it is useful to introduce the conformal time $\eta$ 
by $\eta = {-e^{-Ht}/H} = {-1}/{Ha(\eta)}$, defined in the range 
$-\infty<\eta<0$. The line element is rewritten in a conformally flat form
\begin{eqnarray}
ds^2 = a^2(\eta) (d\eta^2 - d{\bf x}^2).
\label{deline}
\end{eqnarray} 
This spatially flat coordinates cover only half the whole de Sitter spacetime
(The upper left region in Figure \ref{fig:figds}).

By defining the variable scaled by the scale factor $\chi(x) = {\phi(x)}{a(x)}$, 
 we find that the action for $\phi(x)$ is rephrased to the similar form to that 
in the Minkowski spacetime
\begin{eqnarray}
S[\phi] ={1\over 2} \int {d^4 x} {\partial^\mu \chi(x) \partial_\mu \chi(x)}
\end{eqnarray}
with the use of the conformal coordinate and $d^4 x = d\eta d^3 {\bf x}$. 
Then, the two-point function of $\phi_h(x)$ is given by
\begin{eqnarray}
\langle \phi_h(x) \phi_h(y) \rangle = \frac{\langle \chi_h(x) \chi_h(y) \rangle}{a(\eta_x)a(\eta_y)}
= - \frac{1}{4\pi^2 a(\eta_x) a(\eta_y)} \frac{1}{(\eta_x - \eta_y)^2 - |{\bf x} - {\bf y}|^2 - i\epsilon (\eta_x-\eta_y)},\nonumber\\
\end{eqnarray}
and $\langle \phi_h(x) \tilde{\phi}(\omega) \rangle$ becomes
\begin{eqnarray}
\langle \phi_h(x) \tilde{\phi}(\omega) \rangle &=& \int d\tau e^{i\omega \tau} \langle \phi_h(x) \phi_h(z(\tau)) \rangle \nonumber \\ 
&=&  \frac{\eta_x H}{4\pi^2}
\int d\t 
   \frac{e^{i\omega \tau - H \t} }
	     {(\eta_x + e^{-H \t} /H -i\epsilon)^2 - r_x^2 },
\end{eqnarray} 
with $r_x^2 = |{\bf x}|^2$.
The poles of the integrand is obtained by solving the equation
\begin{eqnarray}
(\eta_x + e^{-H \t} /H -i\epsilon)^2 - r_x^2 = 0,
\end{eqnarray}
which yields
\begin{eqnarray}
e^{-H\t}=H(-\eta_x\pm r_x).
\label{eht}
\end{eqnarray}
The structure of the poles on the complex plane of $\tau$ 
depends on a given spacetime point.
The solutions are  
$\t_\pm^R=\xi_{\pm}^R(x)+2\pi n i /H-i\epsilon$ with $n=0,\pm1,\pm2,\cdots$,
for the region $U<0$ and $V>0$, which corresponds to the right Rindler wedge, 
where we defined $r_x + \eta_x  = U$, $r_x - \eta_x = V$, and 
\begin{eqnarray}
  \xi_{\pm}^R(x) = -{1\over H} \ln\left[H (-\eta_x \mp r_x)\right].
\label{detR}
\end{eqnarray}
On the other hand, the solution are 
$\t_-^F=\xi_{-}^F(x)+2\pi n i /H-i\epsilon$ and 
$\t_+^F=\xi_+^F(x)+2\pi (n+1/2) i /H$ with $n=0,\pm1,\pm2,\cdots$, 
for the region $U >0$ and $V >0$, corresponding to the future wedge,
where we defined
\begin{eqnarray}
  \xi_{\pm}^F(x) = -{1\over H} \ln\left[H (\pm \eta_x + r_x)\right].
\label{detF}
\end{eqnarray}

Summing the contributions to the integral, straightforward computations lead to
\begin{eqnarray}
\langle \phi_h(x) \tilde{\phi}(\omega) \rangle = \frac{i}{4\pi a(\eta_x)r_x} 
\frac{1}{e^{2\pi\omega/H}-1}
\left\{ 
e^{i\omega \xi_-(x)} - e^{i\omega \xi^F_+(x)}
e^{\pi\omega/H}\theta(U) - e^{i\omega \xi^R_+(x)}\theta(-U) 
\right\},
\nonumber\\
\end{eqnarray}
which completely agrees with the expression  Eq.(\ref{phitphi}) by replacing $a(\eta_x)r_x$ with 
$l(x)$, and an acceleration constant $a$ with $H$. Similarly, we also have
\begin{eqnarray}
\langle \tilde{\phi}(\omega) \phi_h(x) \rangle = \frac{i}{4\pi a(\eta_x)r_x} 
\frac{e^{2\pi\omega/H}}{e^{2\pi\omega/H}-1}
\left\{ 
e^{i\omega \xi_-(x)} - e^{i\omega \xi^F_+(x)}
e^{-\pi\omega/H}\theta(U) - e^{i\omega \xi^R_+(x)}\theta(-U) 
\right\}.
\nonumber\\
\label{tpp}
\end{eqnarray}
Thus, the two-point function has a similar structure to the Minkowski case. 
Namely, the key relation Eq~(\ref{iyz}) is satisfied in the region $U<0$ and $V>0$, 
but an extra term appears in the region $U>0$ and $V>0$. 
Radiation is cancelled in both regions, and the remaining terms damp rapidly at large distance, 
$a(\eta_x)r_x\rightarrow \infty$,
as will be demonstrated below. Therefore, the remaining terms 
in the energy-momentum tensor are regarded as a polarization cloud. 

\subsubsection{$U<0$ and $V>0$}
The region corresponds to the right Rindler wedge in the Minkowski spacetime.
In the region $U<0$ and $V>0$, the key relation Eq.~(\ref{iyz}) is satisfied with $\rho(\omega)=1/(1-e^{-2\pi\omega/H})$. 
In our model in the de Sitter spacetime, we have
\begin{eqnarray}
 &&G_R(x,y)
= {1\over a(\eta_x)a(\eta_y)} {\delta(\eta_x-\eta_y-|{\bf x}-{\bf y}|)\over 4\pi |{\bf x}-{\bf y}|}
\label{exGR}
\end{eqnarray}
and
\begin{eqnarray}
&& \langle \phi_h(x) \phi_{inh}(y) + \phi_{inh}(x) \phi_h(y) \rangle \nonumber \\
&&=\lambda^2\int d\tau \int {d\omega\over 2\pi} e^{-i\omega \tau} h(\omega)
\Bigl(
 G_R(y,z(\tau)) \langle \phi_{h}(x)\tilde\phi(\omega)\rangle+
 G_R(x,z(\tau)) \langle \tilde\phi(\omega)\phi_{h}(y)\rangle
\Bigr)
\nonumber\\
\label{termA}
\end{eqnarray}
from Eq.~(\ref{inhh}). 
Substituting Eqs.~(\ref{tpp}) and (\ref{exGR}) into (\ref{termA}), and using 
\begin{eqnarray}
\frac{i}{4\pi a(\eta_x)r_x} 
e^{i\omega \t_-(x)} =
{i}
\int d\tau'G_R(x,z(\tau'))e^{i\omega\tau'},
\nonumber
\end{eqnarray}
we find that the inhomogeneous term cancels and that the remaining terms 
in the renormalized two-point function are given by 
\begin{eqnarray}
&& \langle \phi(x) \phi(y) - \phi_h(x) \phi_h(y) \rangle 
= -\frac{i \lambda^2}{(4\pi)^2 a(\eta_x)r_x a(\eta_y)r_y} 
\nonumber \\
&&~~~~~~~~~~~~~~~~
\times
\int {d\omega\over 2\pi} 
\frac{e^{2\pi\omega/H}}{e^{2\pi\omega/H}-1}
\left\{
h(\omega) e^{-i\omega(\xi^R_- (x) - \xi^R_+(y))} -
h(-\omega) e^{-i\omega(\xi^R_+ (x) - \xi^R_-(y))}   \right\} . \nonumber \\
\label{deC}
\end{eqnarray}
From (\ref{exGR}), the retarded Green 
function with the two points on the detector-trajectory is written as
\begin{eqnarray}
G_R(z(\tau),z(\tau'))={\delta(\tau-\tau')\over 4\pi |\tau-\tau'|},\nonumber
\end{eqnarray}
and we have (see also \cite{Murata:2001}), 
\begin{eqnarray}
{\rm Im}{\tilde G}_R(\omega)={\omega \over 4\pi}. 
\end{eqnarray}
Then, the expression of $h(\omega)$ and its pole are the same as those
in the Minkowski spacetime with the Unruh detector. 
Then the poles of $h(\omega)$ in Eq.~(\ref{deC}) give  terms  proportional to $\theta(\xi^R_-(x)-\xi^R_+(y))$
or $\theta(\xi^R_-(y)-\xi^R_+(x))$, which vanish in the coincidence limit. 
The poles from ${e^{2\pi\omega/H}-1}=0$, $\omega=inH$ with an integer $n$
 give a  term decreasing 
rapidly at large distance, and hence do not produce radiation. 

\subsubsection{$U>0$ and $V>0$}
The region corresponds to the future wedge in the Minkowski spacetime.
In this case, the remaining terms in the renormalized two-point function are given by 
\begin{eqnarray}
&& \langle \phi(x) \phi(y) - \phi_h(x) \phi_h(y) \rangle 
= - \frac{i \lambda^2}{(4\pi)^2 a(\eta_x)r_x a(\eta_y)r_y} 
\nonumber \\
&&~~~~~~~~~~~~~~~~
\times
\int {d\omega\over 2\pi} 
\frac{e^{\pi\omega/H}}{e^{2\pi\omega/H}-1}
\left\{
h(\omega)e^{-i\omega(\xi^F_- (y) - \xi^F_+(x))}   
-h(-\omega)e^{-i\omega(\xi^F_+ (y) - \xi^F_-(x))} 
\right\} .
\nonumber \\
\end{eqnarray}
Similar to the above case with $U<0$, the poles of $h(\omega)$ yield terms  proportional 
to $\theta(\xi^F_- (y) - \xi^F_+(x))$ or $\theta(\xi^F_-(x)-\xi^F_+ (y) )$, which 
become zero in the coincidence limit.  
The poles from ${e^{2\pi\omega/H}-1}=0$ give  terms decreasing rapidly at 
large distance. 
Therefore there is no radiation in this region, too.

The surface $U=0$, which divides the spatially flat de Sitter spacetime
into the two regions with $U<0$ and $U>0$, corresponds to the 
cosmological horizon of the detector. 
Namely, $U=0$ is equivalent to the relation $r_x a(t_x)=1/H$, 
and the detector can never be influenced by the events outside 
the cosmological horizon classically. 
In the case of an accelerated detector in the Minkowski spacetime, 
the detector cannot be influenced by the events in the future wedge
beyond the Rindler horizon.
This is an analogy between the models in the de Sitter model and
in the Minkowski spacetime.
In the model of the de Sitter spacetime, however, the radiation emitted 
from the detector always propagates from the region $U<0$ 
to the region $U>0$ across the surface $U=0$. 
Namely any radiation exits the cosmological horizon 
of the de Sitter spacetime as the universe expands. 
This may naturally explain the fact that the radiation in the region
$U>0$ vanishes when the radiation in the region $U<0$ does. 

\section{Conclusions and Discussions} 
In this paper, we investigated the mechanism of  
cancellation of radiation from an Unruh detector coupled with 
a scalar field.
We found that Eq.~(\ref{iyz})  plays a key roll in the cancellation of radiation. 
The correlation of the inhomogeneous solution is indeed cancelled
by an interference effect between the inhomogeneous solution and the vacuum 
fluctuation when Eq.~(\ref{iyz}) holds. 
Since the same relation can be derived from the KMS relation in a thermal field theory (see Appendix), 
the cancellation mechanism of the radiation has its origin in the thermal 
nature of the uniformly accelerated observer.
However, the mechanism of the cancellation of radiation in the future wedge is
a bit different. 
This may be related to the fact that 
the uniformly accelerated detector can observe only events in the right Rindler wedge. 
The key relation Eq.~(\ref{iyz}) does not hold there,
and there is an additional contribution to the two-point function.
In spite of such differences, the radiation cancels out also in 
the future wedge.
We confirmed these behaviors explicitly in two examples, 
the Unruh detector in Minkowski spacetime and a detector at rest in de Sitter spacetime.

The cancellation of radiation can be generalized to interacting massive theories at least perturbatively.
In presence of interactions, we can calculate the two-point function perturbatively
by using  Wick contractions. Since we showed rapid damping of the two-point function 
in a free massless theory, each contribution to 
the two-point function in perturbative expansions  damps more rapidly than
the free case 
so that they cannot give radiation at infinity.

The cancellation of radiation in the right Rindler wedge can be naturally
interpreted as a thermal behavior of the detector,
in accordance with the fact that the key relation  Eq.~(\ref{iyz}) holds there.
Indeed it is related to the fact that the Green function is periodic in the
imaginary direction of the detector's proper time.
(In the Appendix, we see a connection of the key relation to the
KMS relation in an ordinary thermal system.)
However, the cancellation of radiation in the future wedge is not straightforward.
It reminds us of the classical example of the radiation from
 a uniformly accelerated charged particle.
We know that if a particle is uniformly accelerated, it emits Larmor 
radiation. But, in the accelerated observer's frame, 
the particle sits at rest but in a constant gravitational field. In this frame, 
 there is no radiation but for the polarization cloud around the charged particle.
These two pictures seem to be contradictory to the equivalence principle.
The resolution to this paradox is given in \cite{Boulware:1979qj}.
It was shown that the radiation exists only in the future wedge
and  there is  no radiation at all in the right Rindler
wedge that the comoving observer can access.
This classical example indicates that the cancellation of radiation in the future
wedge does not always follow the thermal behavior of the
accelerated detector.
In the present case, since we are considering a quantum system,
the interference effect plays an important role.
It will be interesting to  further investigate how
the detector and the ground state of the quantum field
is entangled across the horizon~\cite{Brout:1995rd}.

%

\section*{Acknowledgments} 
We would like to thank W.G. Unruh, H. Kodama, M. Sasaki, T. Tanaka, J. Garriga and S. Ichinose
for useful conversation. 
The research by S.I. is supported in part by Grant-in-Aid for Scientific Research (19540316) from MEXT, Japan.
We are also supported in part by "the Center for the Promotion of Integrated Sciences (CPIS) "  of Sokendai.
K.Y. acknowledges  useful discussions during the YITP Long-term workshop 
YITP-T-12-03 on "Gravity and Cosmology 2012", and the support by Grant-in-Aid for Scientific researcher
of Japanese Ministry of Education, Culture, Sports, Science and Technology (No.~21540270 and No.~21244033).

\appendix 
\section{On Equation (\ref{iyz}) and The KMS Relation} 
\setcounter{equation}{0}
\label{iyzsection}

In this appendix, we show that the same relation as Eq.~(\ref{iyz}) is 
derived from the KMS relation in the thermal field theory. 
The KMS relation in the thermal field theory is written as
\begin{eqnarray}
G^{\pm}_\beta (t,\overrightarrow{x}) = G^{\mp}_\beta(t\pm i\beta,\overrightarrow{x}),
\end{eqnarray}
where we denote $G^{+}_\beta(x,y) = \langle \phi(x) \phi(y) \rangle_\beta$ and 
$G^{-}_\beta(x,y) =  \langle \phi(y) \phi(x) \rangle_\beta$, and $\langle\cdot\rangle_\beta$
denotes the thermal average in the thermal state with the temperature $T=1/\beta$. 
One can derive the KMS relation as follows,
\begin{eqnarray}
\langle \phi(t,\overrightarrow{x}) \phi(t',\overrightarrow{x}') \rangle_\beta &=& tr [e^{-\beta H} \phi(t,\overrightarrow{x}) \phi(t',\overrightarrow{x}')]/tr[e^{-\beta H}] \nonumber \\
&=&
tr [e^{-\beta H} \phi(t,\overrightarrow{x})e^{\beta H} e^{- \beta H} \phi(t',\overrightarrow{x}')]/tr[e^{-\beta H}] \nonumber \\
&=&
tr [e^{-\beta H} \phi(t',\overrightarrow{x}') \phi(t+i\beta,\overrightarrow{x})]/tr[e^{-\beta H}] \nonumber \\
&=&
\langle \phi(t',\overrightarrow{x}') \phi(t+i\beta,\overrightarrow{x}) \rangle_\beta.
\end{eqnarray}
From this relation, it is easy to show that
\begin{eqnarray}
 \langle \tilde{\phi}(\omega)\phi(x) \rangle_\beta =
 e^{\beta \omega} 
\langle \phi(x) \tilde{\phi}(\omega) \rangle_\beta,
\end{eqnarray}
which is equivalent to 
\begin{eqnarray}
\langle \tilde{\phi}(\omega)\phi(x) \rangle_\beta
={1\over 1-e^{-\beta \omega}}
\langle [\tilde{\phi}(\omega),\phi(x)] \rangle_\beta.
\end{eqnarray}
In this derivation, the periodicity in the direction of the imaginary time is important.

In both examples in section 3 of the present paper, the Green function is periodic 
in the imaginary direction of the detector's proper time, 
since, in the Unruh detector in Minkowski spacetime,
the trajectory is written in term of   $e^{\pm a\tau}$, and in the de Sitter space,
the conformal time is written as $\eta=-e^{-Ht}/H$.
These periodic behaviors are similar to the above thermal property, but contrary to the KMS relation,
the ordering of the fields are not interchanged. So the periodicity itself does not lead to
 the KMS-like relation  (\ref{iyz}). 



\end{document}